\documentclass[12pt]{article}
\usepackage{amssymb}
\usepackage{epsfig}
\usepackage{graphicx}

\newcommand{\av}{\mbox{\boldmath $a$}}

\newcommand{\bb}{\mbox{\boldmath $b$}}
\newcommand{\ca}{\mbox{\boldmath $c$}}

\newcommand\fr{\displaystyle\frac}

\newcommand{\htts}{\mbox{\boldmath$\hat{t}\kern1pt$}}

\newcommand{\qm}{quantum mechanics}
\newcommand{\eref}[1]{Eq. (\ref{#1})}
\begin{document}
\begin{center}

{\bf A Model of Violation and Superviolation of Bell's Inequality in Local Quantum Mechanics}

{Vladimir K. Ignatovich} \\

{Frank Laboratory for Neutron Physics, Joint Institute for Nuclear Research, Dubna, Russia 141980}

\end{center}

\begin{abstract}
A model for violation
and even superviolation of the Bell's inequalities in coincidence
experiments with photons in local \qm\ is presented. The model is based on assumption that
time retardation or losses in an analyzer depend on angle between linear polarization of
an incident photon and the analyzer's axis.
\end{abstract}

%\keywords{EPR paradox; entangled states; Bell's inequalities; uncertainty relations}

Entangled states of separated particles, introduced by~\cite{epr} and~\cite{ba} make \qm\ nonlocal.
This nonlocality is manifested by violation of Bell's inequalities certified in many experiments
where correlation of particle's polarizations in coincidence experiments were measured.
However below it will be shown that violation of the Bell's inequality in coincidence experiments can be
obtained even in the local quantum mechanics.

\begin{figure}[b!]
{\par\centering\resizebox*{12cm}{!}{\includegraphics{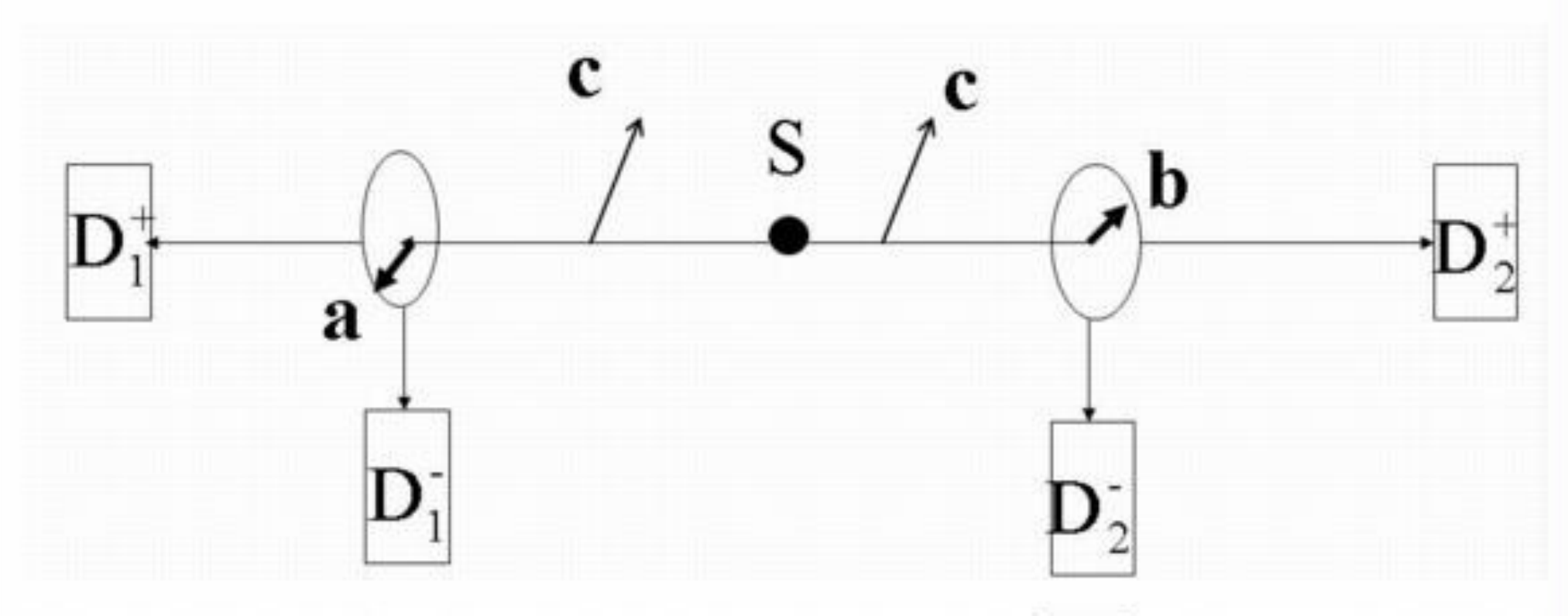}}\par}
\caption{Scheme of the experiment on coincident measurement of a
correlation of polarizations of two photons radiated by the source
$S$. The source radiates two photons with parallel polarizations
$\ca$ which has a uniform angular distribution around direction of
the photons flight paths. Polarizing beam splitters with axes
$\av$ and $\bb$ transmit photons along one of two channels toward
the detectors $D^\pm_{1,2}$.} \label{f1}
\end{figure}

Let's consider an experiment, which scheme is shown in fig.\ref{f1}.
The source $S$ radiates photons assumed to have individual
polarizations along some vector $\ca$, which has uniform angular distribution in the plane orthogonal
to propagation direction. It is the common belief that the Bell's inequalities are not violated
in such a case.
However, it is shown below that they can be violated, which supports
the results of the numerical experiment reported in~\cite{jap}, and more over, in some experiments
one can predict even superviolation, where correlation coefficient surpasses the maximal value $2\sqrt2$.

To be specific let's look at the most popular inequality~\cite{chsh}
\begin{equation}\label{11c0}
-2\le S\le2,
\end{equation}
where
\begin{equation}\label{11c}
S=E(\av,\bb)-E(\av,\bb')+ E(\av',\bb')+E(\av',\bb),
\end{equation}
and $E(\av,\bb)$ is a correlation of polarizations of two
particles registered after two analyzers with their axes along
unit vectors $\av$ and $\bb$ in an experiment depicted in fig.
\ref{f1}.

The definition of correlation function is the most important part of this letter.
Usual definitions involve some predetermined classical functions~\cite{b1} and does not
address the specific features of coincidence experiments such as
arrival time of particles and time window, which means that the width $w$ of the time window is
large enough or $w=\infty$, and no other particle can enter any of
the detectors inside this window.

We suppose that the radiated photons with their individual
polarizations interact with analyzers $\av$ and $\bb$ quantum
mechanically, i.e. probability of a photon with polarization $\ca$
to be transmitted through an analyzer with its axis $\av$ is equal
to $P_+(\av)=(\av\cdot\ca)^2=\cos^2(\alpha-\xi)$, where $\alpha$,
$\xi$ are azimuthal angles of vectors $\av$ and $\ca$ defined with
respect to some axis normal to the propagation direction. In the
following this axis will be chosen along the vector $\av$, so
$\alpha=0$. The angle $\xi$ will be assumed to have uniform
distribution $d\xi/(2\pi)$. Thus the correlation of registrations
looks as in~\cite{jap}
\begin{equation}\label{j}
E(\av,\bb)\equiv
E(\beta)=\fr{P_{++}(\av,\bb)+P_{--}(\av,\bb)-P_{+-}(\av,\bb)-P_{-+}(\av,\bb)}{P_{++}(\av,\bb)+P_{--}(\av,\bb)+P_{+-}(\av,\bb)+P_{-+}(\av,\bb)},
\end{equation}
where, say, $P_\pm(\av,\bb)$ is the probability of registration by
detectors $D_{1,2}^\pm$ in coincidence, and $\beta$ is the angle
between vectors $\av$ and $\bb$.

The analyzers are supposed here to be without losses, and efficiency of
registration after analyzers is supposed to be the same for all
the detectors. Because of definition \eref{j} this efficiency can
be put to unity.
\begin{figure}[t!]
{\par\centering\resizebox*{10cm}{!}{\includegraphics{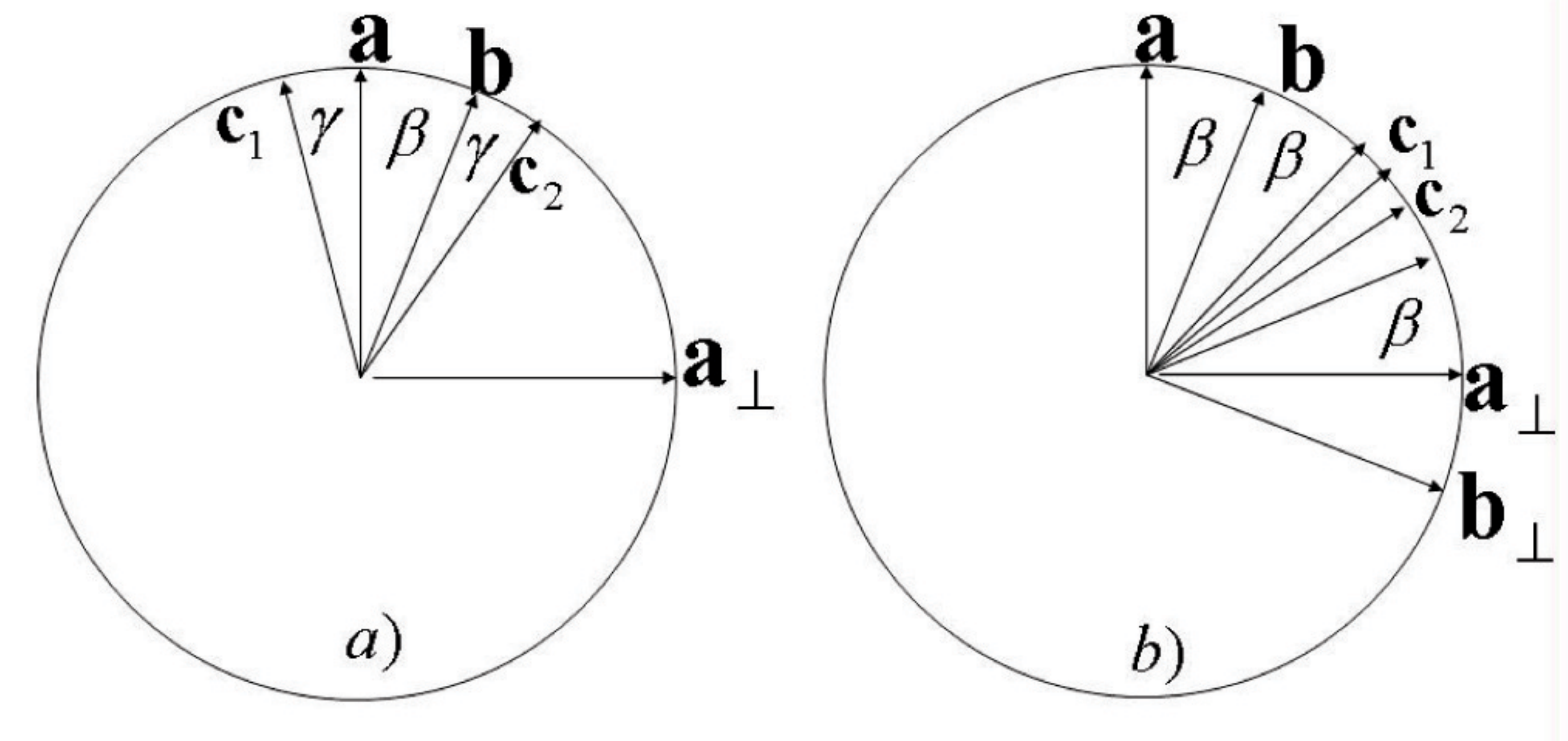}}\par}
\caption{\label{f2}Restriction of integration interval
$(\ca_1,\ca_2)$ in the case $\beta=\beta_0$. Coincidence takes
place only if the angular distance of the photon polarization
$\ca$ from one of the analyzer's axes is not larger than
$\beta+\gamma$ for some fixed $\gamma$. a) Calculation of
probability of the type $P_{++}(\beta=\beta_0)$,
$P_{--}(\beta=\beta_0)$ and also $P_{+-}(\beta=3\beta_0)$,
$P_{-+}(\beta=3\beta_0)$. b) Calculation of probability of the
type $P_{+-}(\beta=\beta_0)$, $P_{-+}(\beta=\beta_0)$ and also
$P_{++}(\beta=3\beta_0)$, $P_{--}(\beta=3\beta_0)$. }
\end{figure}

The probabilities in \eref{j} can be calculated analytically. For
instance,
\begin{equation}\label{j1}
P_{++}(\av,\bb)\equiv P_{++}(\beta)=\int
\fr{d\xi}{2\pi}\cos^2(\xi)\cos^2(\beta-\xi)\Theta(|t_1-t_2|<w),
\end{equation}
where $w$ is the width of the coincidence window, $t_{1,2}$ are
the time delays of the moment of registration and $\Theta$ is the
step function equal to unity, when inequality in its argument is
satisfied, and to zero otherwise.

The goal is to calculate all these probabilities and to show for
some particular case, $\beta=\beta_0=\pi/8$, that the inequality
\begin{equation}\label{1j1}
S=3E(\beta)-E(3\beta)<2,
\end{equation}
can be violated notwithstanding that the photons are not
entangled.

In the following, like in~\cite{jap}, it is supposed that the time
difference $\Delta t=|t_1-t_2|$ depends on angular distance
between vector of photons polarization $\ca$ and axes of
analyzers. For instance, one can suggest that coincidence counting
stops, when this angular distance is larger than $\beta+\gamma$
for some fixed parameter $\gamma$. It means that \eref{j1} can be
represented as
\begin{equation}\label{j2}
P(\beta)=\int\limits_{\xi_1}^{\xi_2}
\fr{d\xi}{2\pi}\cos^2(\xi)\cos^2(\beta-\xi),
\end{equation}
where $\xi_{1.2}$ correspond to limiting positions $\ca_{1,2}$ of
photon polarizations in fig.\ref{f2}. For $P_{++}(\beta)$
integration interval is $-\gamma<\xi<\beta+\gamma$, as is shown in
fig.\ref{f2}a), and for $P_{-+}(\beta)$ integration interval is
$2\beta-\gamma<\xi<\beta+\gamma$, as is shown in fig.\ref{f2}b).
If $P_{++}(\beta)=P_{--}(\beta)$ for $\beta=\beta_0$ and a given
$\gamma$ is denoted as $A(\gamma)$, and
$P_{+-}(\beta)=P_{-+}(\beta)$ is denoted as $B(\gamma)$, then
correlation \eref{j} takes the form
\begin{figure}[t!]
{\par\centering\resizebox*{8cm}{!}{\includegraphics{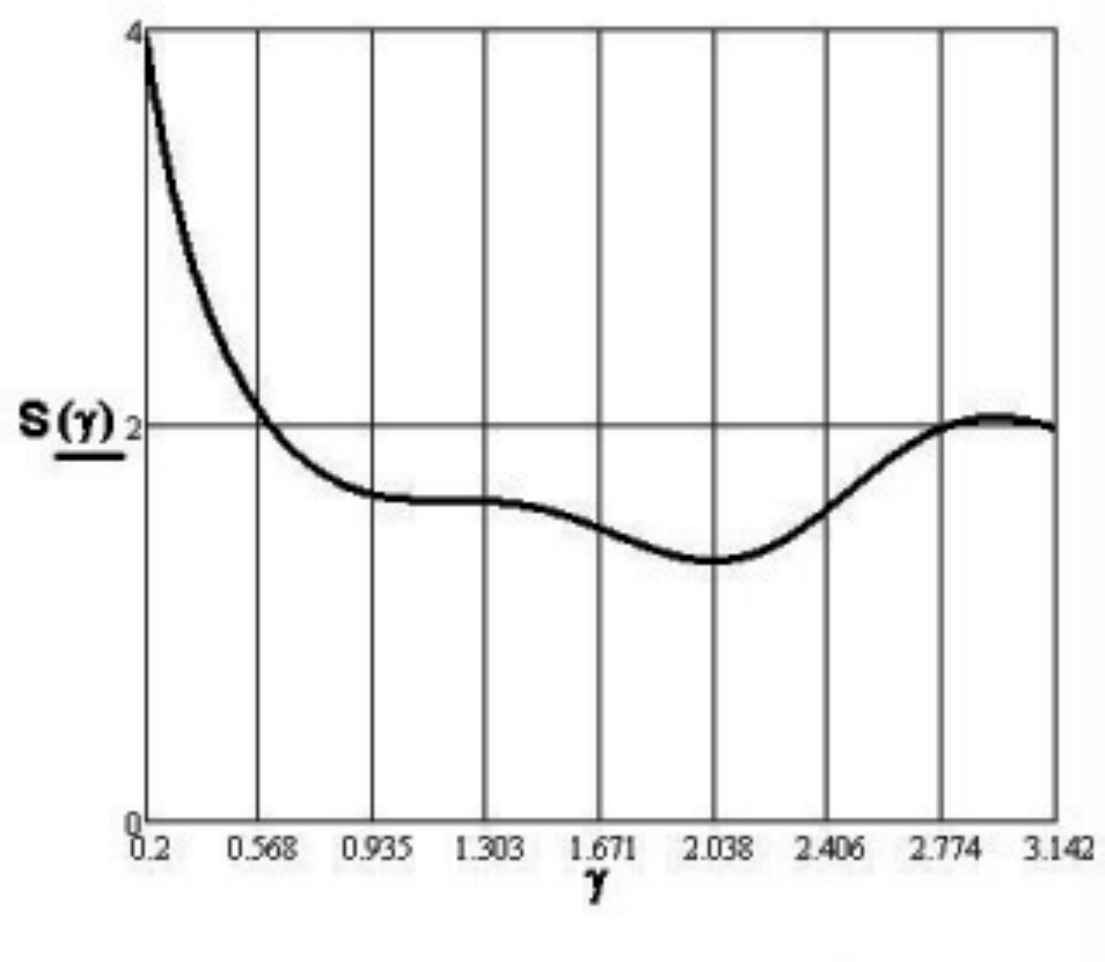}}\par}
\caption{Dependence of $S(\beta_0,\gamma)$ (the first variable is
omitted) on $\gamma$.} \label{f3}
\end{figure}
\begin{equation}\label{j3}
E(\beta_0,\gamma)=\fr{A(\gamma)-B(\gamma)}{A(\gamma)+B(\gamma)},
\end{equation}
and the quantity \eref{1j1}, as is easy to check, becomes
\begin{equation}\label{1j2}
S(\beta_0,\gamma)=3E(\beta_0,\gamma)-E(3\beta_0,\gamma)=4\fr{A(\gamma)-B(\gamma)}{A(\gamma)+B(\gamma)}.
\end{equation}
It is understandable that in the limiting case, when
$\gamma<\beta/2$, the integration interval in fig.\ref{f2}b)
shrinks to zero, therefore $B(\gamma\le\beta_0/2)$ becomes zero
too, and $S(\beta_0,\gamma)$ in \eref{1j2} becomes 4, which is
larger than maximal possible value for entangled states
$S_{\max}=2\sqrt2=2.82$. So in this case we have superviolation of
the Bell's inequality.

Dependence of the function $S(\beta_0,\gamma)$ on $\gamma$, where
the first variable is omitted, is shown in fig.\ref{f3}. It is
seen from there that $S(\gamma)>2$ up to
$\gamma=0.6035\approx1.6\beta_0$.

For conclusion it is necessary to tell that all coincidence experiments devoted to
verification of violation of Bell's inequalities have a preparation stage, when coincidence is tuned.
This preparation stage is never reported. However this stage can be crucial for
``success'' of the experiments, as follows from the above considerations.

\section*{Acknowledgement}
I am to mention here with gratitude the arXiv admins Dr.Jake
Weiskoff and Dr. Don Beyer. Their opinion that my articles on EPR paradox are not much
related to quantum physics, but have a wider significance for all
the general physics is appreciated.


\begin{thebibliography}{9}
\bibitem{epr}
A. Einstein, B. Podolsky, and N. Rosen, {\it Phys. Rev.} , {\bf
47} (1935) 777.
\bibitem{ba}
D. Bohm and Y.Aharonov., Discussion of experimental proof for the
paradox Einstein, Rosen, and Podolsky {\it Phys. Rev.} {\bf108}
(1957) 1070.
\bibitem{jap}
Hans De Raedt, Koen De Raedt, Kristel Michielsen, Koenraad
Keimpema, and Seiji Miyashita. ``Event-based computer simulation
model of Aspect-type experiments strictly satisfying Einstein's
locality conditions.``J. Phys. Soc. Jpn. 76, 104005 (2007);
arXiv:0712.2565v1 [quant-ph].
\bibitem{chsh}
J. F. Clauser, M. A. Horn, A. Shimony, R. A. Holt, Proposed
experiment to test local hidden-variable theories, {\it Phys.Rev.
Lett.} {\bf23} (1969) 880.
\bibitem{b1}
John S. Bell. On the EPR paradox. Physics {\bf 1};195-200:1964.
\end{thebibliography}
\end{document}